\documentclass[12pt,a4paper,english,superscriptaddress,aps,nofootinbib]{revtex4}
\usepackage[utf8]{inputenc}
\usepackage[T1]{fontenc}
\usepackage{amsmath,amssymb,graphicx}
\makeatletter
\usepackage{babel}
\usepackage[active]{srcltx}
\usepackage{graphicx,color}
\usepackage{changebar}
\usepackage{hyperref}
\usepackage[T1]{fontenc}
\usepackage{esint}
\usepackage{multirow}
\usepackage{dsfont}
\usepackage{ae}
\usepackage{amsmath}
\usepackage{braket}
\usepackage{mathtools}
\usepackage{slashed}
\usepackage{empheq}
\usepackage{multirow}
\usepackage{graphicx}
\usepackage{amsfonts}
\usepackage{amsmath}
\begin{document}
\title{On the Structure of Black Bounces Sourced by Anisotropic Fluids}

\author{Leandro A. Lessa}
\email{leandrophys@gmail.com}
\affiliation{Programa de Pós-graduação em Física, Universidade Federal do Maranhão, Campus Universitário do Bacanga, São Luís (MA), 65080-805, Brazil}
\author{Gonzalo J. Olmo}
\email{gonzalo.olmo@uv.es}
\affiliation{Departamento de F\'{i}sica Te\`{o}rica and IFIC, Centro Mixto Universitat de Val\`{e}ncia - CSIC. Universitat de Val\`{e}ncia, Burjassot-46100, Valencia, Spain.}
\affiliation{Universidade Federal do Cear\'a (UFC), Departamento de F\'isica,\\ Campus do Pici, Fortaleza - CE, C.P. 6030, 60455-760 - Brazil.}








\begin{abstract}
The field equations of static, spherically symmetric geometries generated by anisotropic fluids is investigated with the aim of better understanding the relation between the matter and the emergence of minimal area throats, like in wormhole and black bounce scenarios. Imposing some simplifying restrictions on the matter, which amounts to considering nonlinear electromagnetic sources, we find analytical expressions that allow one to design the type of sought geometries. We illustrate our analysis with several examples, including an asymmetric, bounded black bounce spacetime which reproduces the standard Reissner-Nordstrom geometry on the outside all the way down to the throat.  
\end{abstract}

\maketitle


\section{Introduction}

The properties and dynamics of strong gravitational systems play a central role in the quest for understanding the laws that govern the physical world. In this regard, the description provided by classical general relativity (GR) together with {\it natural} energy sources that satisfy certain positivity conditions is far from being satisfactory, as it unavoidably leads to the development of singularities \cite{Penrose:1964wq,Penrose:1969pc,Hawking:1973uf,Senovilla:2014gza,Witten:2019qhl}. 
This motivates the exploration of new theoretical domains and ideas in the hope to find more robust frameworks in which the predictability of the physical laws is retained even in the most extreme scenarios. 

It is in this context that new forms of compact astrophysical objects have been proposed to alleviate the difficulties raised by the classical black hole solutions of GR. A family that has attracted significant interest is that introduced by Visser and Simpson in \cite{Simpson:2018tsi} and dubbed as {\it black bounces}, which represents objects able to  interpolate between nonsingular black holes and wormholes, depending on model parameters. The existence of a minimal 2-sphere, or a bounce, in the radial sector hidden behind an event horizon motivates the name {\it black bounce} to refer to these non-singular objects \cite{Lobo:2020ffi,Mazza:2021rgq}. These objects are free from singularities because the radial function exhibits a non-trivial bouncing behavior that keeps curvature invariants bounded but, more importantly, because they allow geodesic congruences to reexpand beyond the throat to ensure geodesic completeness\footnote{Recall that without geodesic completeness observers and information could be created or destroyed, violating in this way  basic tenets of Physics.} \cite{Carballo-Rubio:2019fnb}.

A diverse array of black bounce models has garnered significant attention in recent literature, with a particular focus on the matter sources that support these configurations \cite{Rodrigues:2023vtm, Pereira:2024rtv, Crispim:2024nou, Pereira:2024gsl}. In particular, Ref. \cite{Bronnikov:2021uta} established that the Simpson-Visser metric is sustained by a matter content composed of nonlinear electrodynamics and a phantom scalar field, where the electromagnetic source corresponds to a magnetic charge. Interestingly, these solutions can also be replicated by considering an electrically charged source \cite{Alencar:2024yvh}. Additionally, the relation between these models and modified gravitational dynamics has been examined in \cite{Alencar:2024nxi, Junior:2024vrv, Junior:2024cbb, Atazadeh:2023wdw, Fabris:2023opv}. Their impact on various physical observables, including gravitational lensing \cite{Nascimento:2020ime, Cheng:2021hoc, Islam:2021ful, Ghosh:2022mka}, black hole shadows \cite{Guerrero:2021ues, Guo:2021wid,Olmo:2023lil,Guerrero:2022qkh}, and gravitational wave echo signals \cite{Silva:2024fpn, Ou:2021efv,Yang:2021cvh}, among other phenomena \cite{Franzin:2022iai, Zhang:2022zox, Yang:2022ryf, Javed:2023iih,daSilva:2023jxa}, has also been extensively investigated.


In this work, we want to better understand the properties of such geometries and the nature of the matter sources able to support them assuming GR as the dynamical background. In fact, we are particularly interested in the way the matter sources contribute to shape the area function that characterizes these solutions, as it is the key to keep curvature invariants within finite bounds. We will thus study the field equations of static, spherically symmetric objects bearing in mind that the function that multiplies the spherical sector may be a non-monotonic function of the radial coordinate. As we shall see, it is possible to make reasonable restrictions on the matter sources, which we will regard as effective anisotropic fluids, to facilitate the construction of exact analytical solutions. This will allow us to build several examples of non-singular black bounce solutions not considered yet in the literature \cite{Alencar:2024yvh,Rodrigues:2022mdm}, including a new type of asymmetric bounded universe \cite{Magalhaes:2023xya,Maso-Ferrando:2023nju}. 

The paper is organized as follows. In Section \ref{sec2}, we present the main equations of motion with an anisotropic fluid as the matter source within the framework of General Relativity. In Section \ref{sec3}, we investigate symmetric black bounce solutions. In Section \ref{sec4}, we explore asymmetric black bounce solutions and analyze their key properties. Finally, in Section \ref{conc}, we summarize the main results and discuss future perspectives for this work. We adopt natural units, $c=\hbar=1$, and use the metric signature $(-,+,+,+)$

\section{Equations of motion} \label{sec2}

The field equations of General Relativity (GR) follow from variation of the Einstein-Hilbert action
\begin{equation}\label{ac}
    S = \frac{1}{2\kappa}\int d^{4}x\sqrt{-g}\bigg[R -2\Lambda +\mathcal{L}_m \bigg],
\end{equation}
where $\kappa = 8\pi G$, $\Lambda$ is a cosmological constant, $G$ represents Newton’s gravitational constant, and $\mathcal{L}_m$ is the matter sector. The equations can be written as
\begin{equation}\label{eqq1}
    G_{\mu\nu} + \Lambda g_{\mu\nu} = T_{\mu\nu},
\end{equation}
 where $T_{\mu\nu}$ is the matter stress-energy tensor. In what follows, and for the sake of generality, we will assume that the matter sector is described by an anisotropic fluid, which will allow us to consider several different matter sources within the same framework. In particular, {we will see that scalar fields,  electromagnetic fields, and a combination of both of them in static and spherically symmetric scenarios can be represented as effective anisotropic fluids}. The stress-energy tensor of an anisotropic fluid is given by
\begin{equation} \label{flui}
    T_{\mu\nu} =(\rho+p_{t})u_{\mu}u_{\nu} + p_t g_{\mu\nu} + (p_r-p_t)v_{\mu}v_{\nu}
    \end{equation}
where $\rho$, $p_r$, and $p_t$ represent the energy density, the radial pressure, and the tangential pressure of the fuid, respectively. The vector $u^{\mu}$ represents the four-velocity of the fluid, normalized as $g_{\mu\nu} u^\mu u^\nu=-1$, while $v_{\mu}$ denotes a space-like unit vector normalized as $g_{\mu\nu} v^\mu v^\nu=+1$. 

For spherically symmetric and static configurations, we can parametrize the line element as
\begin{equation} \label{1}
 ds^{2} = - A(r)dt^2 + B(r) dr^2 + \Sigma(r)^2d\Omega^2
 \end{equation}
where $d\Omega^2=d\theta^2 + \sin^2\theta d\phi^2$ is the element of solid angle on a $2$-sphere. According to this line element, the vectors $u^\mu$ and $v^\mu$ take the form
\begin{eqnarray}
    u^\mu=(\frac{1}{\sqrt{A(r)}},0,0,0) \ , \nonumber \\
    v^\mu=(0,\frac{1}{\sqrt{B(r)}},0,0) \ .
\end{eqnarray}
Additionally, raising one index of the stress-energy tensor with the metric, we see that it becomes diagonal and with components ${T^\mu}_\nu=\text{Diag}(-\rho,p_r,p_t,p_t)$. As mentioned above, the stress-energy tensor of a scalar field in static, spherically symmetric configurations can be seen as equivalent to an anisotropic fluid. If, for instance, one considers a scalar field action of the form 
\begin{equation}
    S_\phi\equiv -\frac{1}{2}\int d^4x \sqrt{-g}F(\phi,X) \ ,
\end{equation}
where $X\equiv g^{\alpha\beta}\partial_\alpha \phi \partial_\beta \phi=g^{rr}\phi_r^2$, one finds
\begin{equation}
    {T^\mu}_{\nu}=F_X g^{\mu\alpha}\partial_\alpha \phi \partial_\nu \phi -\frac{F(\phi,X)}{2}\delta^\mu_\nu =\left(\begin{array}{cccc}
    -\frac{F(\phi,X)}{2} & 0 & 0 & 0 \\
   0 & X F_X -\frac{F(\phi,X)}{2} & 0 & 0 \\
   0 & 0 &  -\frac{F(\phi,X)}{2} & 0 \\
    0 & 0 & 0 & -\frac{F(\phi,X)}{2} 
   \end{array}\right)     \ ,
\end{equation}
from which it is easy to see that $\rho=\frac{F(\phi,X)}{2}=p_t$ and $p_r=F_X -{F(\phi,X)}/{2}$. Similarly, for an electromagnetic source with (nonlinear) action of the form 
\begin{equation}
    S_F\equiv \frac{1}{8\pi}\int d^4x \sqrt{-g}\varphi(\phi,Z) \ ,
\end{equation}
with $Z\equiv - F_{\mu\nu}F^{\mu\nu}/2$, the corresponding stress-energy tensor reads as
\begin{equation}\label{eq:TmnNED}
    {T^\mu}_{\nu}=-\frac{1}{4\pi}\left(Z\varphi_Z {F^\mu}_\alpha {F^\alpha}_\nu-\frac{\varphi(Z)}{2}\delta^\mu_\nu\right) =\left(\begin{array}{cccc}
    \varphi-2{Z\varphi_Z} & 0 & 0 & 0 \\
   0 & \varphi-2{Z\varphi_Z} & 0 & 0 \\
   0 & 0 &  \varphi & 0 \\
    0 & 0 & 0 & \varphi 
   \end{array}\right)     \ ,
\end{equation}
from which wee identify $p_r=-\rho =\varphi-2{Z\varphi_Z}$ and $p_t=\varphi$. Obviously, a combination of a scalar field and an electromagnetic field leads to a more general anisotropic fluid, with $\rho={F(\phi,X)}/{2}+\varphi-2{Z\varphi_Z}$, $p_r= X F_X -{F(\phi,X)}/{2}+\varphi-2{Z\varphi_Z}$, and $p_t=\varphi-{F(\phi,X)}/{2}$. Thus an effective anisotropic fluid can cover many matter distributions of interest in a relatively simple way. \\

Due to the freedom in the choice of radial coordinate, it is sometimes useful to use the function $\Sigma(r)$ as a radial coordinate, but given that we are interested in wormhole and black bounce solutions, where the area of the $2$-spheres may have a non-zero lower bound (a minimum), it is convenient to keep $\Sigma(r)$ as an unspecified function of $r$ in such a way that $r$ can act as a valid coordinate even across the minimum of $\Sigma(r)$. If we  used $\Sigma$ as the radial coordinate, the term $B(r) (dr/d\Sigma)^2$ at the minimum would be ill defined because $d\Sigma/dr=0$ at that location. Having this in mind, the field equations (\ref{eqq1}) evaluated on the line element (\ref{1}) become
\begin{equation}\label{eq1}
\frac{B' \Sigma '}{B^2 \Sigma}-\frac{2 \Sigma ''}{B \Sigma}-\frac{\Sigma '^2}{B \Sigma ^2}+\frac{1}{\Sigma ^2} =  \rho, 
\end{equation}
\begin{equation}\label{eq2}
   \frac{A' \Sigma '}{A B \Sigma }+\frac{\Sigma '^2}{B \Sigma ^2}-\frac{1}{\Sigma ^2} = p_r
\end{equation}
and
\begin{equation}\label{eq3}
  \frac{A''}{2 A B}-\frac{A' B'}{4 A B^2}+\frac{A' \Sigma '}{2 A B \Sigma}-\frac{A'^2}{4 A^2 B}-\frac{B' \Sigma '}{2 B^2 \Sigma}+\frac{\Sigma ''}{B\Sigma }=p_t.
\end{equation}
On the other hand, from the conservation of the stress-energy  tensor, $\nabla_{\mu}T^{\mu}{}_{\nu}=0$ one obtains
\begin{equation} \label{conse}
  \frac{A'}{4 A}\bigg( \frac{\rho+p_r}{p_t-p_r} \bigg)+\frac{p_r '}{2(p_t-p_r)}=    \frac{\Sigma '}{\Sigma } \ .
\end{equation}
The above set of four equations involves six unknown functions, namely, $\left\{A, B, \Sigma, \rho, p_r, p_t \right\}$, which demands some additional inputs in order to solve the system. Since our focus here is to understand the mechanisms that determine the dependence of $\Sigma(r)$ on the matter sources, we will impose some restrictions on the matter sector to facilitate the analysis. To motivate the constraints we will use, let us first add Eqs.(\ref{eq1}) and (\ref{eq2}) to obtain
\begin{equation} \label{AB0}
    \frac{1}{B}\bigg[ \bigg(\frac{1}{AB}\frac{d(AB)}{dr}\bigg)\frac{ \Sigma '}{\Sigma} -\frac{2 \Sigma ''}{ \Sigma} \bigg]=\rho+p_r.
\end{equation}

Given the expressions (\ref{conse}) and (\ref{AB0}), we see that important progress can be achieved by considering matter sources for which the combination $\rho+p_r$ vanishes. This happens, in particular, for Maxwell electrodynamics but also for all nonlinear theories of electrodynamics under the assumption of staticity and spherical symmetry, as one can see right below Eq.(\ref{eq:TmnNED}). Restricting our attention to such kind of sources, we find that (\ref{AB0}) becomes 
\begin{equation} \label{AB}
    \frac{1}{B}\bigg[ \bigg(\frac{1}{AB}\frac{d(AB)}{dr}\bigg)\frac{ \Sigma '}{\Sigma} -\frac{2 \Sigma ''}{ \Sigma} \bigg]=0,
\end{equation}
which can be integrated once to yield 
\begin{equation}\label{rel}
    B = \frac{\Sigma '^2}{A} \ ,
\end{equation}
where an irrelevant integration constant has been absorbed in a redefinition of the coordinate $r$. Substituting the relation (\ref{rel}) into the original line element (\ref{1}), it is easy to see that it can be written as
\begin{equation} \label{1m}
 ds^{2} = - A(\Sigma)dt^2 +  \frac{d\Sigma^2}{A(\Sigma)} + \Sigma^2d\Omega^2 \ ,
 \end{equation}
which is a very standard form of writing the line element when $\Sigma$ is used as the radial coordinate. Recall, however, that $\Sigma$ can only be used as a coordinate in domains where $d\Sigma/dr\neq 0$ and, therefore, it is unable to convey all the additional information encoded in the parametrization $\Sigma(r)$. For this reason, one should not forget that (\ref{1}) contains more information about the geometry than (\ref{1m}).

The form of the function $A(\Sigma)$ can then be obtained from Eq. (\ref{eq2}) by assuming that $p_r=-\rho$ can be written as a function of $\Sigma$, which leads to 
\begin{equation}\label{eq:AofSigma}
    \Sigma A_{\Sigma} + A = 1 - \rho \Sigma^2 \ ,
\end{equation}
where we are denoting $A_{\Sigma}= \frac{\partial A}{\partial \Sigma}$. This equation can be further simplified with the ansatz $A=1-2M(\Sigma)/\Sigma$, leading to $M_\Sigma=\rho \Sigma^2/2$, which is the usual Newtonian expression for the mass function and can be directly integrated if a function $\rho=\rho(\Sigma)$ is given. This is what we discuss next.

We can now turn our attention to the conservation equation  (\ref{conse}), which under the assumption $\rho+p_r=0$ turns into 
\begin{equation} \label{conse1}
 \frac{\Sigma '}{\Sigma }=-\frac{\rho'}{2(p_t+\rho)}
\end{equation}
By specifying an equation of state of the form $p_t=p_t(\rho)$, this equation can be integrated by quadratures, yielding a relation of the form 
\begin{equation}
    \Sigma=\Sigma_0 e^{-\int\frac{d\rho}{2(p_t(\rho)+\rho)}} \ ,
\end{equation}
where $\Sigma_0$ is an integration constant. In many cases of interest it will be possible to invert this relation and find an explicit expression for $\rho=\rho(\Sigma)$, which is the ingredient we needed in Eq.(\ref{eq:AofSigma}) to obtain $A(\Sigma)$, as pointed out above. 

So far we have used the conditions $\rho+p_r=0$ and $p_t=p_t(\rho)$ to obtain $A(\Sigma)$ and $\rho(\Sigma)$, but we still need to determine the form of $B$ and of $\Sigma(r)$. To proceed further, we could specify the form of $B=B(\Sigma)$ and then use Eq.(\ref{rel}) to find $\Sigma(r)$ by direct integration. Alternatively, we could specify $\Sigma(r)$ and find $B(r)$ using that relation (\ref{rel}). In this work, we adopt this latter approach. To progress in this direction, we use Eq. (\ref{conse1}) which relates the areal function to the matter content. We know that the minimum condition for the function is given by $\Sigma'(r_m)=0$ and $\Sigma''(r_m)>0$, where $r_m$ sets the location of the  minimum. This implies that the right-hand side (RHS) of Eq. (\ref{conse1}) vanishes. For simplicity, we choose the following parameterization for the RHS
\begin{equation} \label{para}
  \frac{\Sigma '}{\Sigma } = f((r - r_{m})^{n}).  
\end{equation}
where $n$ is an integer. Note that at $r=r_m$ the first minimum condition is already satisfied if $f(0)=0$. To meet the other condition, we need to differentiate both sides of the equation above and then evaluate it at $r=r_m$. Thus, we show that the minimum for the function $\Sigma$ is obtained only if $f'(r_m)>0$ and $n=1$ (for $n>1$ we find that  $\Sigma''(r_m)=0$, whereas for $n<1$ we have $\Sigma''(r_m)\to \infty$ ). Without loss of generality, we can assume this minimum occurs at the origin $r_m=0$. So that the next step is to assume some profiles for the function $f$ respecting the condition $f(0)>0$. Subsequently, we need to designate an equation of state $p_t = p_t(\rho)$ to determine $A(r)$ and $B(r)$.

\section{Symmetric Bounce Models} \label{sec3}
\subsection{$f(r) = \frac{r}{a^2+r^2}$}
Our first choice for the function $f$ is motivated by the Visser-Simpson solution \cite{Simpson:2018tsi}. It is straightforward to show from Eq.(\ref{para}) that if we consider $f(r) = \frac{r}{a^2+r^2}$, the function $\Sigma$ is given by
\begin{equation}\label{vs}
\Sigma=\Sigma_0\sqrt{a^2+r^2},
\end{equation}
where $\Sigma_0$ is an irrelevant integration constant that can be set to unity without loss of generality. The areal radius (\ref{vs}) exhibits the regular minimum $\Sigma(0)=a$, with $a$ taken to be the throat radius. Once the form of this function is determined, we need only to define an equation of state to fully solve the system of Eq. (\ref{eq1},\ref{eq2},\ref{eq3}, \ref{conse}). For simplicity, we adopt that $p_t=\omega \rho$, where $\omega$ is a dimensionless constant. By substituting this choice along with the solution for $\Sigma$ into Eq. (\ref{conse1}), we can obtain the energy density profile for this configuration given by
\begin{equation}\label{energi}
    \rho (r)= \frac{\rho_0}{(a^2+r^2)^{1+\omega}}
\end{equation}
where $\rho_0$ is a constant. 

Now, substituting Eq. (\ref{rel}) into Eq. (\ref{eq1}) with $\Sigma=\sqrt{a^2+r^2}$ and (\ref{energi}), we find that
\begin{equation}
    A(r) = 1 - \frac{2m}{\sqrt{a^2+r^2}} + \frac{\rho_0 }{(2\omega-1)(a^2+r^2)^{\omega}} ,
\end{equation}
where $m$ is a constant related to the mass. And finally, we find through Eq. (\ref{rel}) that
\begin{equation}
      B(r)^{-1} =\bigg(1 + \frac{a^2}{r^2} \bigg)\bigg( 1 - \frac{2m}{\sqrt{a^2+r^2}} + \frac{\rho_0 }{(2\omega-1)(a^2+r^2)^{\omega}} \bigg)
\end{equation}
Note that in the limit $a=0$, we recover Kiselev solution \cite{Kiselev:2002dx}. In the special case where $\omega=1$, we obtain the black bounce form of the Reissner-Nordström solution, with the parameter $\rho_0$ being interpreted as the charge. 

It is important to emphasize that although we used the same function $\Sigma$ from the Visser-Simpson work and found the same charged black bounce solution for the function $A$, our solution is strictly different because $A\neq B$. Additionally, the difference becomes clear when we compare the energy density and pressures in this work with those obtained in \cite{Simpson:2018tsi}.

\begin{figure*}
 \includegraphics[height=5cm]{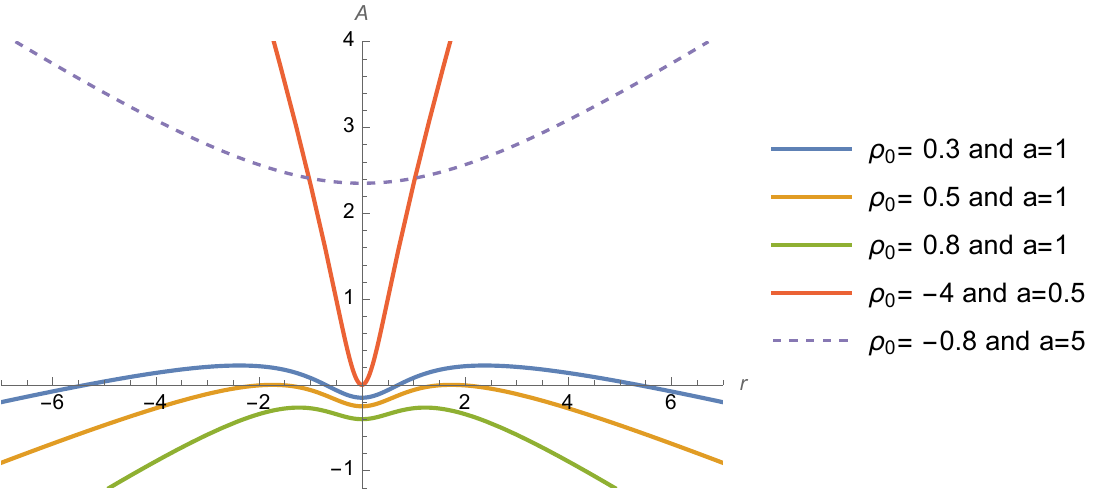}
\caption{The representation of the metric function $A$, obtained from the line element (\ref{vss}), is shown as a function of the radial coordinate $r$ for $m = 0.5$. The throat is located at the origin.}  \label{fig1}
\end{figure*}

Let us now investigate some geometric properties for the case $\omega = -1/2$, a black bounce solution surrounded by quintessence given by
\begin{align}\label{vss}
    ds^2 = -&\bigg(1 - \frac{2m}{\sqrt{a^2+r^2}} - \frac{\rho_0 \sqrt{r^2+a^2} }{2} \bigg)dt^2 + \frac{dr^2}{\bigg(1 + \frac{a^2}{r^2} \bigg)\bigg(1 - \frac{2m}{\sqrt{a^2+r^2}} - \frac{\rho_0 \sqrt{r^2+a^2} }{2} \bigg)}\\\nonumber
    &+ (a^2+r^2)d\Omega^2. 
\end{align}
In the limit $a=0$, we recover the Schwarzschild black hole surrounded by quintessence \cite{Fernando:2012ue}. Calculating the Kretschmann scalar,  
\begin{equation}
    K = R_{\alpha\beta\mu\nu}K^{\alpha\beta\mu\nu} = \frac{2 \left(\rho_0^2 \left(a^2+r^2\right)^2+24 M^2\right)}{\left(a^2+r^2\right)^3}
\end{equation}
we see that the spacetime is indeed regular everywhere. The location of horizons for the line element (\ref{vss}) is determined by $g_{tt}=A(r_h)=0$, which leads to
\begin{equation}
    r_h = S_1\sqrt{\frac{\left(1+S_2\sqrt{1-4 M \rho_0}\right)^2}{\rho_0^2}-a^2}
\end{equation}
where $S_1,S_2=\pm 1$, the positive sign of $S_1$ defines the horizon on our universe, $r>0$, while the negative sign corresponds to the universe on the other side of the throat. On the other hand, $S_2$ corresponds to an outer (positive sign) or inner (negative sign) horizon. This leads to five cases of interest:
\begin{enumerate}
    \item For $4m<\frac{1}{\rho_0}$ and $a<\frac{1+S_2\sqrt{1-4 M \rho_0}}{ \rho_0 }$, we have a regular black hole with a standard outer (inner) horizon (thick blue line in Fig.\ref{fig1}). 
    \item For $\rho_0=\frac{1}{4m}$ and $a\leq4m$, 
we have an extreme black hole with an extremal horizon at $r_{ext}=\pm\sqrt{16m^2-a^2}$ (thick orange line in Fig.\ref{fig1}).
    \item For $4m>\frac{1}{\rho_0}$, we have no horizons in this geometry. This case corresponds to a traversable wormhole. (thick green line in Fig.\ref{fig1})
    \item For $4m<\frac{1}{\rho_0}$ and $a=\frac{1+S_2\sqrt{1-4 M \rho_0}}{ \rho_0 }$, the four horizons converge at $r_h=0$, which can be regarded as a hyperextremal configuration (thick red line in Fig.\ref{fig1}).
    \item For $4m<\frac{1}{\rho_0}$ and $a>\frac{1+S_2\sqrt{1-4 M \rho_0}}{ \rho_0 }$, there are  no horizons, leading to a traversable wormhole (dashed line in Fig.\ref{fig1}).
.
\end{enumerate}

\subsection{$f(r) = \tanh \left(\frac{\alpha \ r}{r_0}\right)$}
Our second example corresponds to  $f(r) = \tanh \left(\frac{\alpha \ r}{r_0}\right)$, for which $\Sigma$ has a minimum if the dimensionless constant \(\alpha\) is positive. Substituting this choice into Eq.(\ref{para}), we find that
\begin{equation}
\Sigma=\Sigma_0\left(\cosh \left[\frac{\alpha r}{r_0}\right]\right)^{\frac{1}{\alpha}}. 
\end{equation}
Here, \(r_0\) is a constant with the dimension of length. Unlike in the previous example, here the constant $\Sigma_0$ is relevant because it determines the size of the throat at $r=0$, $\Sigma(0)=\Sigma_0$. The energy density from Eq.(\ref{conse1}) when  $p_t=\omega \rho$ can be written as
\begin{equation}\label{energ}
   \rho(r)=\frac{\rho_0}{ \left[\cosh\left(\frac{\alpha r}{r_0}\right)\right]^{\frac{2 (\omega +1)}{\alpha }}}=\rho_0\left(\frac{\Sigma_0}{\Sigma(r)}\right)^{2(1+\omega)}
\end{equation}
where $\rho_0$ represents the maximum density at the throat. The line element for this configuration can be written in the form of Eq.(\ref{1m}) with the function $A(\Sigma)$ given by
\begin{equation}\label{eq:ASigma}
    A(\Sigma)=\left\{\begin{array}{lr} 
    1-\frac{2\tilde{m}}{\Sigma(r)}-\frac{\rho_0 \Sigma_0^2}{(1-2\omega)}\left(\frac{\Sigma_0}{\Sigma(r)}\right)^{2\omega} & \text{ if } \omega\neq 1/2 \\  \\
    1-\frac{2\tilde{m}}{\Sigma(r)}-\frac{\rho_0 \Sigma_0^3}{\Sigma(r)}\ln\left(\frac{\Sigma(r)}{\Sigma_0}\right) & \text{ if } \omega= 1/2\end{array}\right. \ ,
\end{equation}
where $\tilde{m}$ is an integration constant that represents the asymptotic ADM mass. 

\begin{figure*}
 \includegraphics[height=5cm]{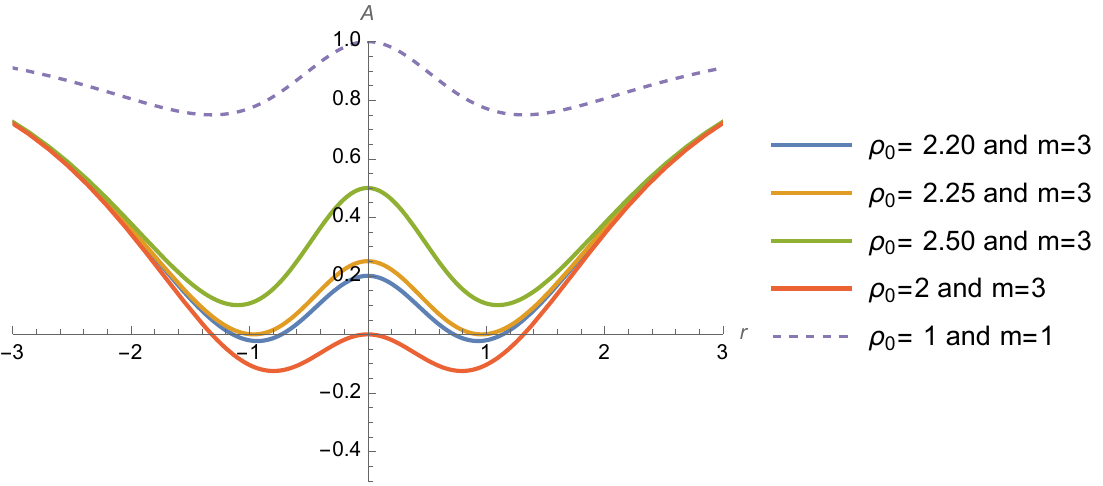}
\caption{The representation of the metric function $A$, obtained from the line element (\ref{ex2}), is shown as a function of the radial coordinate $r$ for $\Sigma_0=1$. The throat is located at the origin.}  \label{fig2}
\end{figure*}

For the sake of clarity, we will now adopt a specific equation of state to further explore the properties of the solution. In this instance, we assume $\omega = 1$, which corresponds to recovering Maxwell-like electrodynamics as the matter content. We will also set $\alpha = r_0 = 1$ for simplicity. Consequently, the line element governed by Maxwell-like electrodynamics with the areal function given by \ref{energ} can be expressed as
\begin{align}\nonumber \label{ex2}
    ds^2 =- &\bigg(1 -m \text{sech}(r)+\Sigma_0^2 \rho_0 \text{sech}^2(r) \bigg)dt^2 + \frac{\sinh(r)dr^2}{ \Sigma_0^2 \bigg(1 -m\text{sech}(r)+\Sigma_0^2 \rho_0 \text{sech}^2(r) \bigg)} \\ 
    &+  \Sigma_0^2 \cosh (r)d\Omega^2 ,
\end{align}
where we are denoting $m\equiv 2\tilde{m}/\Sigma_0$. 
We shall now examine some properties of this solution. Calculating the Kretschmann scalar, we find
\begin{equation}
    K = R_{\alpha\beta\mu\nu}K^{\alpha\beta\mu\nu} = 4 \text{sech}^6(r) \left(\frac{3 m^2}{\Sigma_0^4}+\frac{12 m \rho_0 \text{sech}(r)}{\Sigma_0^2}+14 \rho_0^2 \text{sech}^2(r)\right).
\end{equation}
Once again, we obtain a spacetime that is regular everywhere. Moreover, the location of the horizon for the solution is given by
\begin{equation}
    r_h = S_1 \cosh ^{-1}\bigg(\frac{m+S_2\sqrt{m^2-4 \rho_0 \Sigma_0^2}}{2}\bigg)
\end{equation}
where again $S_1,S_2=\pm 1$.
\begin{enumerate}
    \item For $m^2>4 \rho_0 \Sigma_0^2$ and $\frac{m+S_2\sqrt{m^2-4 \rho_0 \Sigma_0^2}}{2}>1$, we have a regular black hole with a standard outer (inner) horizon (thick blue line in Fig.\ref{fig2}). 
    \item For $m^2=4 \rho_0 \Sigma_0^2$ and $m\geq 2$, 
we have an extreme black hole with an extremal horizon at $r_{ext}=\pm \cosh ^{-1}( m/2)$ (thick orange line in Fig.\ref{fig2}).
    \item For $m^2<4 \rho_0 \Sigma_0^2$, we have no horizons in this geometry. This case corresponds to a traversable wormhole (thick green line in Fig.\ref{fig2}).
      \item For $m^2>4 \rho_0 \Sigma_0^2$ and $\frac{m+S_2\sqrt{m^2-4 \rho_0 \Sigma_0^2}}{2}=1$, This case corresponds to a non-traversable wormhole since the geometry possesses an event horizon at the throat (thick red line in Fig.\ref{fig2}).
    \item For $m^2>4 \rho_0 \Sigma_0^2$ and $\frac{m+S_2\sqrt{m^2-4 \rho_0 \Sigma_0^2}}{2}<1$, the geometry has no inner or outer horizons, becoming a traversable wormhole (dashed line in Fig.\ref{fig2}).
\end{enumerate}

\section{Asymmetric bounded bounce model}\label{sec4}
\subsection{$f(r) = \frac{\alpha^2}{r}\bigg(1-\frac{r_0}{r}\bigg)e^{-\frac{l}{r}}$}

\begin{figure*}
 \includegraphics[height=5cm]{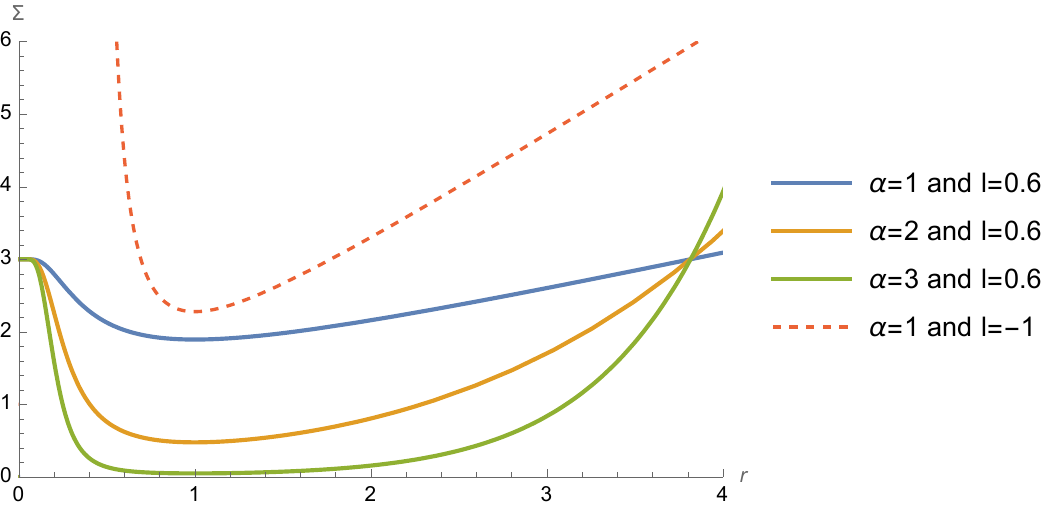}
\caption{Representation of the function $\Sigma(r)$ defined in Eq.(\ref{asyy}) for $\tilde{\Sigma}_0=3$. The throat is located at $r_0=1$. Note that the area of the maximal sphere at $r=0$ for $l>0$, $A_0=4\pi\Sigma_0^2$ can be much larger than the area at the minimum $r=r_0$, which rapidly goes to zero as the parameter $\alpha$ grows. Note that the curves with $l>0$ converge at a point $r>r_0$ where the function $\chi(r)$ vanishes.  }  \label{fig3}
\end{figure*}

As a final example, we explore an asymmetric areal function defined by $f(r) = \frac{\alpha^2}{r}(1-{r_0}/{r})e^{-\frac{l}{r}}$, where $\alpha, r_0$, and $l$ are constants, and $r$ is defined only in the interval $r\ge 0$.   Substituting this choice into Eq.(\ref{para}), we find that
\begin{equation}\label{asyy}
\Sigma=\tilde{\Sigma}_0 e^{-{\alpha^2}\chi(r)} \ ,
\end{equation}
where we have defined 
\begin{equation}\label{chi}
    \chi(r)\equiv\text{Ei}\left(-\frac{l}{r}\right)+\frac{r_0}{l} e^{-\frac{l}{r}}
\end{equation}
and $\text{Ei}(z)$ represents the exponential integral function, $\text{Ei}(z)=-\int_{-z}^{\infty}\frac{e^{-t}}{t} dt$. 
The minimum of the areal radius is located at $r=r_0$, as can be seen in  Fig.\ref{fig3}. An important property of the function $\chi$ is that for $l>0$ we have that $\chi\to0$ as $r\to0^{+}$, which implies that $\tilde{\Sigma}_0$ is the value at $r\to0^{+}$, not at the minimum $r=r_0$, where $\Sigma_0\equiv \Sigma(r_0)<\tilde{\Sigma}_0$. 
This type of configuration is analogous to the {\it bounded universes} recently introduced in \cite{Magalhaes:2023xya}
to illustrate that a minimum in the radial sector needs not represent a wormhole-like structure and that more exotic objects are still possible. In Fig.\ref{fig3} we can see that after crossing the throat at $r=r_0$, one enters into a region bounded by a maximal sphere of area $4\pi \tilde{\Sigma}_0^2$ located at $r\to 0$. Note that had we chosen $l < 0$, the resulting structure would not be bounded. Instead, the sign in front of the second term in (\ref{chi}) changes and the exponential diverges as $r\to 0$, leading to a minimum at $r=r_0$ followed by an unbouded growth, which provides an asymmetric bounce. This is also illustrated in the dashed curve shown in  Fig. \ref{fig3}.

For the above radial function and considering $p_t=\omega \rho$, we see that the energy density from Eq.(\ref{conse}) can be written in the standard form 
\begin{equation}\label{energ}
\rho(r)=\tilde{\rho}_0\left(\frac{\tilde{\Sigma}_0}{\Sigma(r)}\right)^{2(1+\omega)} \ ,
\end{equation}
where now $\tilde{\rho}_0$ denotes the energy density at $r\to 0^+$ for $l>0$. For negative $l$, it is easy to see that the energy density, as given by Eq. \ref{energ}, vanishes for $\omega>-1$ at $r\to 0^+$. 
As in the previous example, the expression for the metric function $A(\Sigma)$ is formally the same as in Eq.(\ref{eq:ASigma}) but with the parametrization given in Eq.(\ref{asyy}), namely, 
\begin{equation}\label{eq:ASigma2}
    A(\Sigma)=\left\{\begin{array}{lr} 
    1-\frac{2\tilde{m}}{\Sigma(r)}-\frac{\tilde{\rho}_0 \tilde{\Sigma}_0^2}{(1-2\omega)}\left(\frac{\tilde{\Sigma}_0}{\Sigma(r)}\right)^{2\omega} & \text{ if } \omega\neq 1/2 \\  \\
    1-\frac{2\tilde{m}}{\Sigma(r)}-\frac{\tilde{\rho}_0 \tilde{\Sigma}_0^3}{\Sigma(r)}\ln\left(\frac{\Sigma(r)}{\tilde{\Sigma}_0}\right) & \text{ if } \omega= 1/2\end{array}\right. \ ,
\end{equation}
For the discussion of horizons, it is convenient to use the variable $\Sigma$ instead of $r$ because it facilitates the qualitative understanding of the possible configurations. For concreteness, we will consider the case $\omega=1$, for which we get a kind of Reissner-Nordstrom configuration with 
\begin{equation}\label{as}
    A(\Sigma)=1-\frac{2\tilde m}{\Sigma}+\frac{\tilde{\rho}_0 \tilde{\Sigma}_0^4}{\Sigma^2} \ ,
\end{equation}
where $\rho_0 \tilde{\Sigma}_0^4$ plays the role of an effective charge squared. The zeros of this function determine the number and location of horizons, if any. Since $\Sigma(r)$ is not single-valued in an interval around $r_0$ and one can find up to two zeros in the function $A(\Sigma)$, denoted as $\Sigma_{\pm}$, the total number of horizons on the $r$-axis will depend on the particular interval on which those zeros appear. 

\begin{figure*}
 \includegraphics[height=7cm]{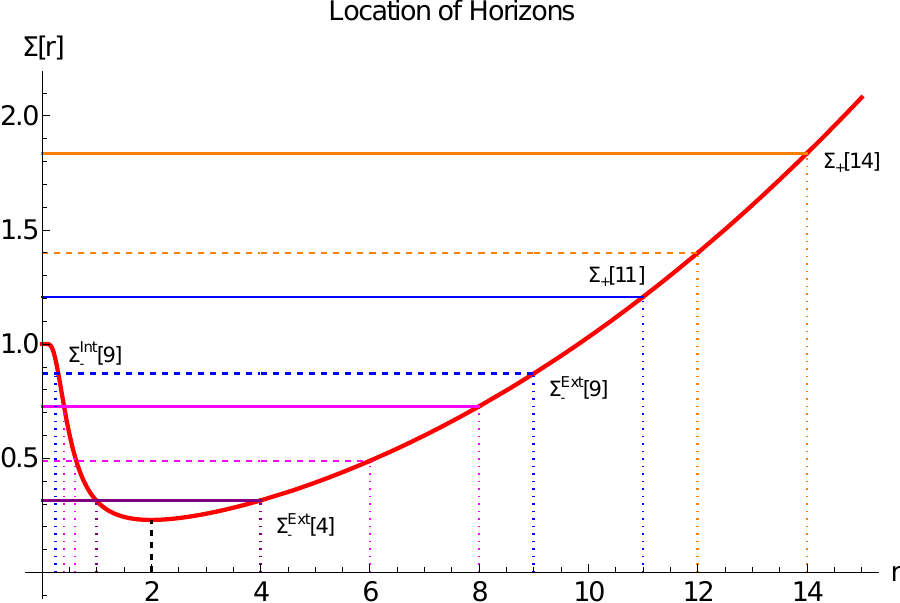}
\caption{Qualitative illustration of different horizon configurations depending on the value of $\Sigma(r)$. The solid and dashed orange curves intersect the red curve $\Sigma(r)$ only at one location, representing a typical pair of (inner and outer) horizons of the Reissner-Nordstrom geometry, with the solid/dashed line being the outer/inner horizon. The blue curves represent a hybrid situation, where the outer horizon only has one intersection, denoted as $\Sigma_+[11]$, while the inner one has two, denoted as $\Sigma_-^{Int}[9]$ and $\Sigma_-^{Ext}[9]$, respectively.}  \label{HorizonsM3}
\end{figure*}

Considering first the case $l>0$, if $\Sigma_{\pm}>\tilde{\Sigma}_0$, then we will only have two horizons, with the throat ($r=r_0$) and the maximum internal sphere ($r=0$) located below the inner horizon ($r_0<r_-$). If $\Sigma_0<\Sigma_{\pm}<\tilde{\Sigma}_0$ then we will have two horizons before the bounce ($r>r_0$) and other two in the region $r<r_0$ (between $\Sigma_0$ and $\tilde{\Sigma}_0$). If $\Sigma_{+}$ is smaller than $\Sigma_0$, then there will be no horizons at all. If $\Sigma_-<\Sigma_0$ and $\tilde{\Sigma}_0>\Sigma_+$, then there will be two $\Sigma_+$ horizons, one on each side of $\Sigma_0$.     If $\Sigma_-<\Sigma_0$ and $\tilde{\Sigma}_0<\Sigma_+$, then there will only be one $\Sigma_+$ horizon in the region $r>r_0$. See Fig.\ref{HorizonsM3} for an illustration of the above cases.

One can check that the Kretschmann scalar for this solution takes the form
\begin{equation}\label{eq:KretRN}
     K = R_{\alpha\beta\mu\nu}R^{\alpha\beta\mu\nu} = \frac{8\left(7\tilde{\rho}_0^2\tilde{\Sigma}_0^8-12 \tilde{m} \tilde{\rho}_0\tilde{\Sigma}_0^4\Sigma(r)+6 \tilde{m}^2 \Sigma(r)^2\right)}{\Sigma(r)^8} \ .
\end{equation}
The function $\Sigma(r)$ never vanishes, as for $r_0 \neq 0$, $\chi$ does not diverge to positive infinity for any value of $l$. Consequently, curvature invariants, such as the Kretschmann scalar (\ref{eq:KretRN}) , remain regular throughout this geometry. Furthermore, we find that the Kretschmann scalar (\ref{eq:KretRN}) vanishes as $r \to 0^{+}$ for $l < 0$, since $\chi \to -\infty$, i.e., $\Sigma(0^{+})\to \infty$ if $l<0.$


Returning to the radial coordinate $r$ in Eq. (\ref{as}), we now explore various possible gravitational objects. Notably, as $r \to 0^{+}$ for $l>0$, we observe that $\chi \to 0$, implying that $A(r)$ remains regular at the origin and approaches a constant value:
\begin{equation}
    A(0^{+}) \to1-\frac{2\tilde{m}}{\Sigma_0}+\rho_0 \Sigma_0^2 \ \ \  \text{if} \ \ \  l>0 \ .
\end{equation}
In Fig. \ref{fig4}, we present examples of the three distinct types of gravitational objects that follow from the above expression. The blue curve corresponds to a traversable wormhole with its throat located at $r = 0.5$. The orange curve represents an extremal black hole, while the green curve illustrates a standard black hole featuring two horizons. Notably, in both the extremal and non-extremal black hole cases, a throat persists below their respective  horizons at $r = 0.5$, with the maximum internal spherical surface located at $r = 0$.

{To conclude this section, it remains to discuss the solution for $l < 0$. In this case, since $\chi \to -\infty$ as $r \to 0^{+}$, we have that $\Sigma \to \infty$ in this region and leads to 
\begin{equation}
    A(0^{+}) \to 1 \ \ \  \text{if} \ \ \  l<0 \ .
\end{equation}
Therefore, the internal region represents an asymptotically Minkowskian space-time. This if further verified by noting that  the energy density and curvature scalars like the Kretschmann scalar vanish.
 }

\begin{figure*}
\includegraphics[height=5cm]{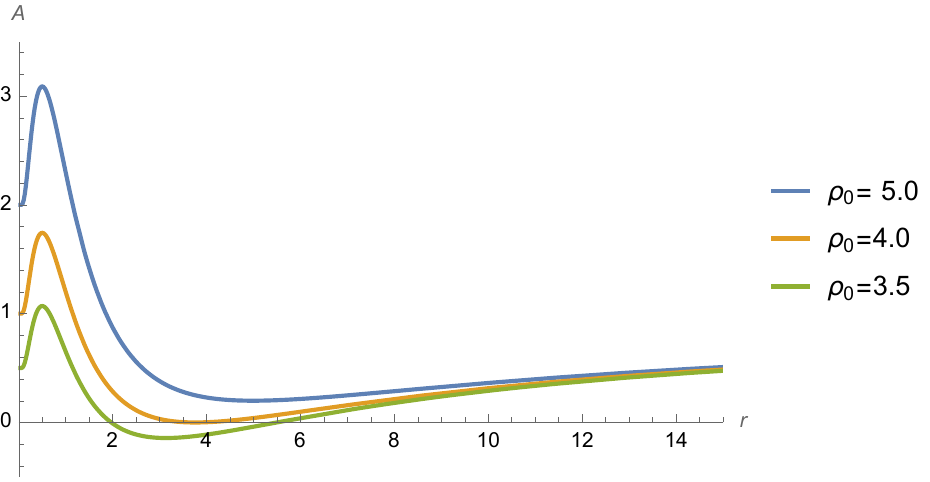}
\caption{The representation of the metric function $A$ defined in (\ref{as}), in terms of the radial coordinate $r$ for $\tilde{\Sigma}_0 = \alpha = 1$ and $l = 0.5$ (positive). The throat is located at $r_0 = 0.5$. Notably, for a mass $\Tilde{m} = 2$, we encounter different gravitational objects for varying values of $\rho_0$.
}  \label{fig4}
\label{fig4}
\end{figure*}

\section{Summary and conclusion}\label{conc} 

In this work we have studied the relation between the area function of spherically symmetric black bounces and the matter sources within the framework of GR. Our aim was to better understand how the matter fields shape this function in order to be able to design new compact objects geometries of physical interest. After deriving the general equations that govern such geometries with an anisotropic fluid as the matter source, we have imposed some restrictions on the matter sector to facilitate the derivation of analytical expressions. These restrictions amount to considering nonlinear electrodynamics sources, for which a general formula for the metric function $A(\Sigma)$ can be found. Further progress was possible by setting an equation of state that relates the tangential pressure with the energy density (see Eq.(\ref{conse1})), and from that relation, the conditions to impose a bouncing geometry were clear, which paved the way to propose several illustrative examples. In particular, we have presented two symmetric black bounce families of solutions and an asymmetric one (bounded if $l>0$ or unbounded if $l<0$). The symmetric examples include a Kiselev-like solution modified à la Visser-Simpson, with the replacement $r\to \sqrt{r^2+a^2}$, and another case in which the area function is parametrized as a hyperbolic cosinus. The regularity and horizons structure has been discussed for concrete values of the equation of state. For the asymmetric example, we have considered a new type of exotic object that locally develops a minimal area surface, like usual wormholes and black bounces, but which globally is topologically trivial (when $l>0$) due to the fact that after crossing the throat the universe becomes bounded, like a kind of space-time bubble attached to a throat. For simplicity, we have chosen the equation of state in such a way that the solution on the unbounded side of the throat looks like the usual Reissner-Nordstrom black hole. However, the presence of the throat together with the existence of a maximal sphere on the other side (two length scales that compete with the size of the horizons) give rise to a richer structure of horizons.

The fact that the objects presented here are completely regular offers an opportunity to study the phenomenology of geometries that from a distance may look like naked singularities but whose innermost structure cures their associated pathologies.
The thermodynamics and phenomenological implications of internal structures of these compact objects will be explored in a forthcoming publication following the approach of \cite{Magalhaes:2023xya}.  More general mechanisms to generate bouncing geometries with sources beyond nonlinear electrodynamics will also be considered.

\acknowledgements{The authors express their gratitude to FAPEMA and CNPq (Brazilian research agencies) for their invaluable financial support. L.A.L is supported by FAPEMA BPD- 08975/24. The authors also acknowledge financial support from the Spanish Grant i-COOP23096 funded by CSIC, from PID2020-116567GB-C21, PID2023-149560NB-C21 funded by MCIN/AEI/10.13039/501100011033, and by CEX2023-001292-S funded by MCIU/AEI.  The paper is based upon work from COST Action CaLISTA CA21109 supported by COST (European Cooperation in Science and Technology).
}

\end{document}